\documentclass[preprint,nofootinbib]{revtex4}%
\usepackage{amssymb}
\usepackage{amsfonts}
\usepackage{amsmath}
\usepackage{amsmath}
\usepackage{graphicx}
\usepackage[usenames]{color}%
\providecommand{\U}[1]{\protect\rule{.1in}{.1in}}
\providecommand{\U}[1]{\protect\rule{.1in}{.1in}}
\definecolor{blue}{rgb}{0,0,1}

\definecolor{red}{rgb}{1,0,0}

\begin{document}
\title{$R^{2}$ corrections to the black string instability and the boosted black string}
\author{Carla Henríquez-Báez$^{1,2}$, Julio Oliva$^3$, Marcelo Oyarzo$^3$, Marcel I. Yáñez Reyes$^3$}
 
\affiliation{$^1$ Departamento de Matemática y Física Aplicadas, Universidad Católica de la Santísima Concepción, Concepción, Chile}
\affiliation{$^2$ 
Instituto de Ciencias Físicas y Matemáticas, Universidad Austral de Chile,
Casilla 567, Valdivia, Chile}
\affiliation{$^3$Departamento de F\'{\i}sica, Universidad de Concepci\'on, Casilla, 160-C, Concepci\'on, Chile.}

\begin{abstract}
We consider a perturbative Gauss-Bonnet term supplementing the Einstein-Hilbert action, and evaluate its effect on the spectrum of the scalar mode that triggers the Gregory-Laflamme instability of black strings in five dimensional General Relativity. After studying some properties of the static black string, we provide the correction to the Lichnerowicz operator up to $\mathcal{O}(\alpha^2)$. For the scalar mode of the gravitational perturbation, we find a master variable and study its spectrum, providing an analysis of the regime of validity of our scheme. We show that the instability persists under the inclusion of the $R^2$-correction, and that the critical wavelength increases with the value of $\alpha/r_+^2<<1$, providing new evidence of the phenomenon which was already observed in other approaches.  We also construct the boosted black strings and compute the correction to the mass, the momentum and the tension due to the higher curvature term. The presence of the dimensionful coupling $\alpha$ spoils the validity of the Smarr relation, which is the gravitational version of the Euler relation which must hold for every homogeneous, thermodynamic system. We show that this identity can be restored by working in an extended thermodynamic setup that includes variations of the Gauss-Bonnet coupling.  

\end{abstract}
\maketitle

\section{Introduction}
Black strings and black branes can be defined as black hole spacetimes with
horizons that have extended directions, which in the simplest case are planar \cite{Horowitz-book}. These spacetimes have a very interesting dynamics,
since in general they suffer from the Gregory-Laflamme (GL) instability \cite{GL1, GL2}
, which is triggered by a gravitational perturbation travelling along
an extended direction with a wavelength above a given critical value. In a
remarkable series of works \cite{Lehner-Pretorius-1, Lehner-Pretorius-2, Lehner-Pretorius-3}, the
authors were able to find strong numerical evidence in favour of the
pinching-off of the horizons of black strings in a finite time. This phenomenon is compatible
with the previous no-go results by Horowitz and Maeda \cite{Horowitz-Maeda},
since it refers to a finite value of the time for asymptotic observers,
providing an example of violation of cosmic censorship in dimension five for
generic initial data. This result has
been recently confirmed in \cite{Figueras-Andrade-reciente}, where the authors were able to
numerically evolve the spacetime closer to the pinch-off, and provide evidence
of a non-geometric progression for the time intervals of the appearance of new
generations of black holes connected by black strings. The non-linear
evolution of the system was also addressed in the context of the Large-D
expansion of General Relativity (GR) \cite{Emparan-Suzuki-Tanabe-1, Emparan-Suzuki-Tanabe-2, Emparan-Suzuki-Tanabe-3}, giving rise
to a non-uniform black string as the final configuration after the GL
instability is triggered, which is consistent with the existence of a critical
dimension obtained by Sorkin in \cite{Sorkin-criticalD}. In references \cite{Wiseman-numerics-black-string, Gubser-NUBS}
non-uniform black strings were constructed numerically and perturbatively, and in \cite{Figueras-Kun-1, Figueras-Kun-2, Figueras-Kun-3, Andrade-Fig} new numerical simulations of the fully non-linear Einstein equations provide evidence of violations of cosmic censorship triggered by GL instabilities for asymptotically Minkowski spacetimes.

As the black string evolves, regions with higher curvature will be exposed,
and it is natural to expect that higher curvature corrections to gravity may
play a role. Remarkably, in the recent paper \cite{Reall-Kovacs-Stronghyperbolicity} a new gauge was found for
initial value problem in Einstein-Gauss-Bonnet gravity, which leads to a
strongly hyperbolic system for bounded curvatures. This may allow to evolve
the black string in the presence of higher curvature terms, and study their
effect on the dynamics of the system.

In this paper we study the black string instability spectrum of backgrounds
that are corrected at leading order in the Gauss-Bonnet parameter. This
precise $R^{2}$ correction can be obtained from string theory as an expansion
on the string tension. In such a framework, in order to construct
solutions and study their stability beyond linear order in $\alpha$ in a
consistent manner, one would have to consider higher powers of the curvature as
corrections, therefore if one insists in interpreting our results as string
theory corrections to the GL instability one can not go beyond linear terms in
$\alpha$, having always in mind that one is indeed working in an effective
field theory setup. As mentioned below, this also affects the regime of
validity of the gravitational perturbation. We also construct the boosted
black string of this theory and obtain the first order corrections in $\alpha$
to the energy, momentum, entropy, temperature and tension.

\section{The corrected, static, black string}
We will consider the Einstein-Gauss-Bonnet action \cite{Lovelock-original}

\begin{equation}
I\left[  g\right]  =\int d^{5}x\sqrt{-g}\left(  R+\alpha\left(  R_{ABCD}%
R^{ABCD}-4R_{AB}R^{AB}+R^{2}\right)  \right)  +O\left(  \alpha^{2}\right)
\ .\label{action}%
\end{equation}
as an effective field theory, to first order in the coupling $\alpha$, which has mass dimension $-2$.

We start by re-obtaining the closed form static black string solutions, which were originally obtained in 
\cite{Brihaye-Delsate-Radu-BS-EGB} to first order in $\alpha$. Additionally, we provide  thermodynamic quantities associated to this spacetime.

Let us consider a black string ansatz in dimension five, in the field
equations of the theory defined by (\ref{action}):
\begin{equation}\label{metricstring}
ds^{2}=-f\left(  r\right)  dt^{2}+\frac{dr^{2}}{g\left(  r\right)  }%
+r^{2}d\sigma^2+b\left(  r\right)
dz^{2}\ ,
\end{equation}
where $d\sigma$ is the line element of a two-sphere. We assume that the metric functions $X=\left\{  f,g,b\right\}  $ are analytic in $\alpha=0$, therefore they can be expanded as $X=X_{0}+\alpha
X_{1}+\mathcal{O}\left(  \alpha^{2}\right)  $. The introduction of a dimensionful scale
$\alpha$ forces us to consider a non-constant warp factor, $b\left(
r\right)  $ \cite{Giribet-Oliva-Troncoso, Kastor-Mann}. Dropping terms $\mathcal{O}\left(  \alpha^{2}\right)  $, the
system of equations for the functions $X_{0}$ and $X_{1}$ can be integrated in
a closed manner. The general solution involves four new integration constants
on top the integration constant of the $X_{0}$ functions, namely the mass
parameter of the GR solution, as well as logarithmic terms in the radial
coordinate. Using the freedom under coordinate transformations, the
perturbative scheme in $\alpha$ and requiring the $\alpha$-corrected spacetime
to have an event horizon leads to the following expressions in terms of the
radius of the horizon $r_{+}$:%
\begin{align}\label{bs5df}
f\left(  r\right)   &=f_5\left(  r\right)  =1-\frac{r_{+}}{r}-\frac{\left(  r-r_{+}\right)
\left(  6r_{+}^{2}+11rr_{+}+23r^{2}\right)  }{9r_{+}r^{4}}\alpha+O\left(
\alpha^{2}\right)  \ ,\\\label{bs5dg}
g\left(  r\right)   & =g_5\left(  r\right) =1-\frac{r_{+}}{r}+\frac{\left(  r-r_{+}\right)
\left(  r+5r_{+}\right)  \left(  r+2r_{+}\right)  }{9r_{+}r^{4}}%
\alpha+O\left(  \alpha^{2}\right)  \ ,\\\label{bs5db}
b\left(  r\right)   &= b_5\left(  r\right) =1+\frac{4\left(  6r^{2}+3rr_{+}+2r_{+}^{2}\right)
}{9r_{+}r^{3}}\alpha+O\left(  \alpha^{2}\right)  \ ,
\end{align}
The temperature, mass and entropy of this black string are respectively given by
\begin{align}
T&=\frac{1}{4\pi r_+}-\frac{11}{36\pi r_+^3}\alpha+\mathcal{O}\left(\alpha^2\right)  \ , \\
M&=8\pi r_{+}L_{z}+\frac{88\pi}{9r_{+}}L_{z}\alpha+\mathcal{O}\left(  \alpha^{2}\right)\ , \\
S&=16 \pi^2 r_+^2 L_z+\frac{928 \pi^2}{9}L_z \alpha+\mathcal{O}\left(
\alpha^{2}\right) \, 
\end{align}
where the former was computed from the surface gravity in Eddington-Finkelstein-like coordinates, and the latter two were computed using the Iyer-Wald method \cite{Iyer-Wald-1, Iyer-Wald-2}. These expressions fulfil the first law of black hole thermodynamics
\begin{equation}
dM=TdS\ ,
\end{equation}
disregarding quadratic terms in $\alpha$. Here, $L_z$ is the length of the extended direction with coordinate $z$. Notice that both the correction to the mass and entropy are positive.

\section{The perturbation}
The generalized Lichnerowicz operator, namely, the linearized field equations
around a generic background $\mathring{g}_{\mu\nu}$, read
\begin{eqnarray}
&&0=\frac{1}{2}\left( -\delta _{\mu }^{\rho }\delta _{\nu }^{\lambda
}\mathring\square +2\mathring{R}_{\ \mu \nu }^{\rho \ \ \lambda }+\mathring{g_{\mu \nu }}\mathring{R}^{\rho \lambda
}+4\mathring{g}^{\lambda \eta }\mathring{G}_{\eta (\mu }\delta _{\nu )}^{\rho }\right) h_{\rho
\lambda }-2\alpha \mathring{R}_{\mu \xi \lambda \nu }\mathring{\square} h^{\lambda \xi }
\notag \\
&&+\alpha \left( -\mathring{R}_{\ \ \ \ \ \lambda }^{\rho \eta \sigma }%
\mathring{R}_{\xi \sigma \rho \eta }\mathring{g}_{\mu \nu }-2\mathring{R}_{\
\ \ \lambda \nu }^{\sigma \rho }\mathring{R}_{\sigma \rho \xi \mu }+4%
\mathring{R}_{\mu \ \ \nu }^{\ \rho \sigma }\mathring{R}_{\lambda \rho
\sigma \xi }+4\mathring{R}_{\mu \rho \lambda \sigma }\mathring{R}_{\nu \ \ \
\xi }^{\ \sigma \rho }\right) h^{\lambda \xi }  \notag \\
&&+\alpha \left( 4\mathring{g}_{\lambda (\mu }\mathring{R}_{\nu )\rho \sigma
\xi }\mathring{\nabla}^{\sigma }\mathring{\nabla}^{\rho }h^{\lambda \xi }+4%
\mathring{\nabla}^{\sigma }\mathring{\nabla}_{(\mu }h^{\xi \lambda }%
\mathring{R}_{\nu )\xi \lambda \sigma }-2\mathring{g}_{\mu \nu }\mathring{R}%
_{\ \lambda \xi \eta }^{\rho }\mathring{\nabla}_{\rho }\mathring{\nabla}%
^{\eta }h^{\lambda \xi }\right)+\mathcal{O}\left(\alpha^2\right) \ ,  \notag
\end{eqnarray}
where we have imposed transversality and tracelessness of the perturbation,
i.e. $\mathring\nabla_{\mu}h^{\mu\nu}=0$ and $h_{\mu}^{\ \ \mu}=0$. We have also used the vacuum field equations, which allow to write the Ricci tensor and Ricci scalar in terms of an expression that is linear in $\alpha$, that can be used in the linearization of the Gauss-Bonnet tensor to write every Ricci tensor and Ricci scalar in terms of the Riemann tensor plus $\mathcal{O}\left(\alpha^2\right)$ terms.

\textbf{On the regime of validity of the perturbation:}
The black string is parameterized by the coordinates $(t,r,x^i,z)$, where the $x^i$ collectively denote the coordinates on a round sphere. We will focus on the s-wave, scalar mode, which is the responsible for the GL instability in GR. Therefore the metric perturbation that we are considering reads
\begin{equation}
h_{AB}=\varepsilon e^{\Omega t}e^{ikz}\left(
\begin{array}
[c]{cccc}%
H_{tt}\left(  r\right)   & H_{tr}\left(  r\right)   & 0 & 0\\
H_{tr}\left(  r\right)   & H_{rr}\left(  r\right)   & 0 & 0\\
0 & 0 & H\left(  r\right)  \sigma_{ij} & 0\\
0 & 0 & 0 & 0
\end{array}
\right)\sim e^{\Omega t}e^{i k z}H_{\mu\nu}(r)  \ ,\label{perturbationstring}%
\end{equation}
with $\varepsilon<<1$ and where $\sigma_{ij}$ denotes the metric on the sphere.

Schematically, when evaluated on the perturbation, the bulk Lagrangian will have the form
\begin{align}
R+\alpha R^2
&\sim \text{Background}+\partial h \partial h +\alpha \left(\partial h\partial h\right)^2+\mathcal{O}\left(\alpha^2\right) \ .
\end{align}
Due to the separation in modes, the terms $\partial h \partial h +\alpha \left(\partial h\partial h\right)^2$ will contain the following contributions
\begin{align}
T_1&=k^2+\alpha k^4+ \mathcal{O}\left(\alpha^2\right)\ , \\
T_2&=\Omega^2+\alpha \Omega^4 +\mathcal{O}\left(\alpha^2\right)\ ,
\end{align}
from the derivatives with respect to $z$ and to $t$, respectively. In consequence, in order to ensure the validity of the perturbative approach, we impose
\begin{align}
\alpha k^2<<1\ \text{and} \ \alpha \Omega^2<<1\ ,
\end{align}
namely, a sufficient condition for the validity of dropping-off terms $\mathcal{O}\left(\alpha^2\right)$, is to keep attention on the modes with small $k$ and small $\Omega$ as compared with $\alpha^{-1/2}$.
We are interested in the existence and behavior of unstable modes, therefore,
for a given $k$, after imposing the boundary conditions for the perturbations,
we look for positive values of $\Omega$ that may allow to connect the regular
asymptotic behavior of the perturbation both at the horizon and infinity.

As in GR, the dynamics of the scalar mode, defined by the functions $\left\{  H_{tt},H_{tr},H_{rr},H\right\}
$ is completely controlled by the master variable $H_{tr}\left(  r\right)  $,
and the remaining functions are given in terms of the master variable and its
derivatives. The second order, linear, homogeneous ODE for $H_{tr}\left(
r\right)  $ has the form%
\begin{equation}
A\left(  r;\alpha\right)  \frac{d^{2}H_{tr}}{dr^{2}}+B\left(
r;\alpha\right)  \frac{dH_{tr}}{dr}+C\left(  r;\alpha\right)
H_{tr}=0\ ,\label{ecuHtr}%
\end{equation}
where the coefficients depend on $r$ and are linear in $\alpha$, which is consistent
with the perturbative approach we are considering. We do not provide the
explicit expression for the coefficients since they are not illuminating.

The asymptotic expansions of (\ref{ecuHtr}) near the horizon and infinity lead to%
\begin{align}
H_{tr}\left(  r\right)  &\sim\left(  r-r_{+}\right)  ^{-1\pm\Omega\left(
r_{+}+\frac{11\alpha}{9r_{+}}+\mathcal{O}\left(  \alpha^{2}\right)  \right)  }\left(  1+O\left(r-r_+\right)  \right)  \ ,
\\
H_{tr}\left(  r\right)   &  \sim\frac{e^{\pm r \sqrt{\Omega^{2}+k^{2}}}}%
{r^{\xi_{\pm}}}\left(  1+O\left(r^{-1}\right)  \right),
\end{align}
where $\xi_{\pm}$ are constants which are not-relevant for recognizing the
regular asymptotic behavior, dominated by the exponential growing/suppression. In consequence, we choose the $\left(  +\right)
$-branches in the near horizon expansion and the $\left(  -\right)  $-branches as $r$ goes to infinity. 

In order to find the
spectrum, we proceed as follows: we introduce the function $F\left(  r\right)
$ such that
\begin{align}
H_{tr}\left(  r\right)  &=\left(  r-r_{+}\right)  ^{-1+\Omega\left(
r_{+}+\frac{11\alpha}{9r_{+}}+\mathcal{O}\left(  \alpha^{2}\right)  \right)  }F\left(
r\right)  \ ,
\end{align}
and rewrite the equation for $F\left(  r\right)  $ in terms of the coordinate
$x$ such that $r=r_{+}/\left(  1-x\right)  $ which maps $r\in]r_{+},+\infty
\lbrack$ to $x\in]0,1[$. Then, we select the regular branch at the horizon by
assuming that $F\left(  x\right)  $ has a Taylor expansion around $x=0$. We
therefore define the truncated series as%
\begin{equation}\label{truncated}
F_{N}\left(  x\right)  =1+\sum_{j=1}^{N}a_{j}\left(  k,r_{+},\Omega, \alpha\right)
x^{j}\ ,
\end{equation}
and solve the equation, near the horizon ($x\rightarrow0$) for the
coefficients $a_{j}$. For a given wavenumber $k$, horizon radius $r_{+}$ and
value of the coupling $\alpha$, the frequencies are obtained by setting
$F_{N}\left(1\right)=0$, for a large enough $N$, such that a notion of
numerical stability for the frequency $\Omega$ is attained. Actually, one can introduce the dimensionless quantities $\hat{\Omega}=r_+\Omega$ and $\hat{k}=r_+ k$ and observe that \eqref{truncated} depends only on the pair $(\hat{k},\hat{\Omega})$ and the dimensionless ratio $\alpha/r_+^2$. We have also used shooting to validate our numerical results.
\begin{figure}[htb]
\begin{center}
\includegraphics[scale=0.8]{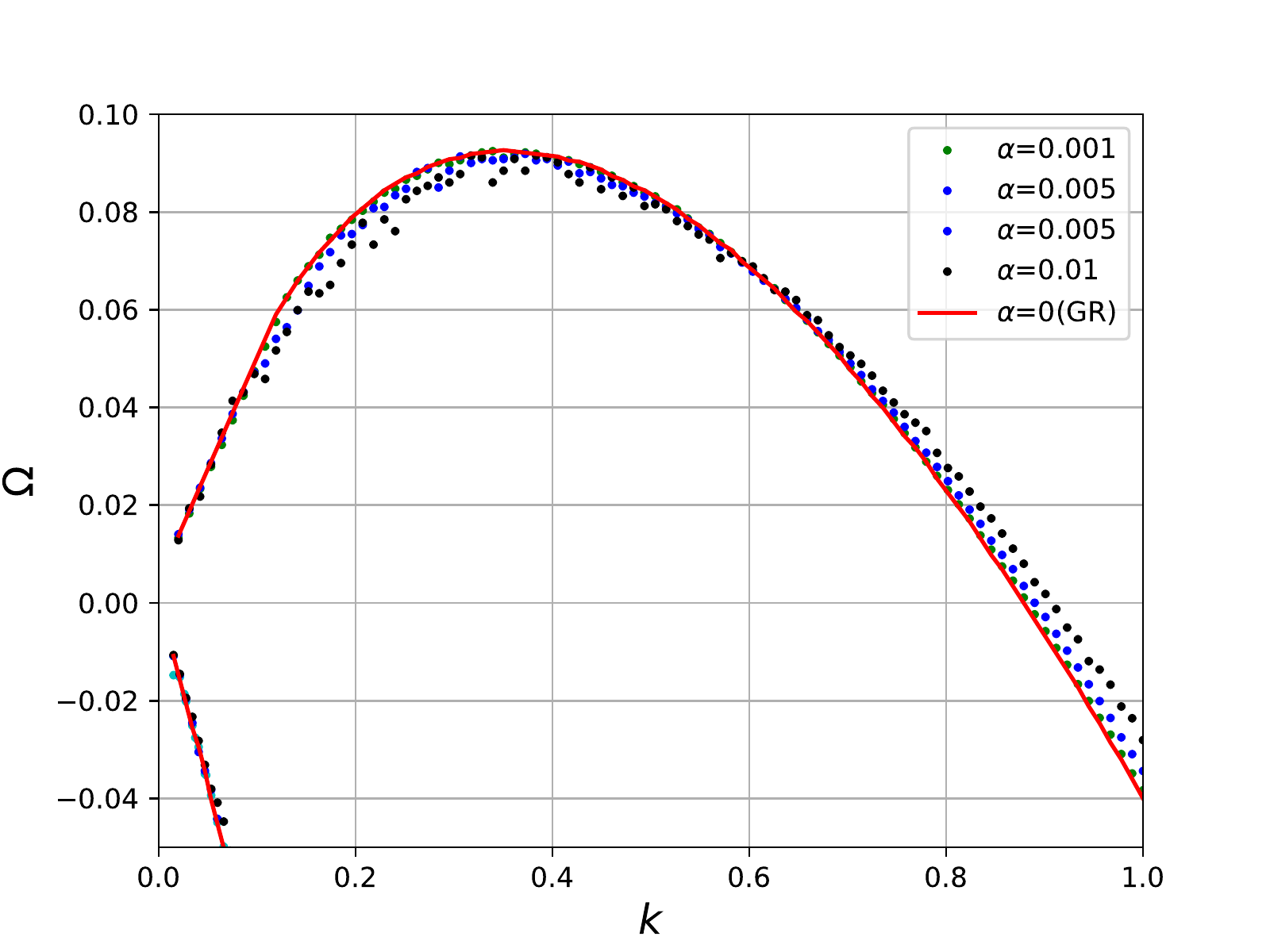}
\caption{Spectra of the perturbation for different values of the $\alpha$ correction, and $r_+$=1.}
\end{center}
\end{figure}
Figure 1 depicts the spectrum of the scalar perturbation in GR (red lines), as well as its correction for different values of the coupling $\alpha$. Aside from the unstable modes, with positive $\Omega$, we also depict a second stable mode with $\Omega<0$, which is present in GR and corrected due to the presence of the perturbative Gauss-Bonnet term. Our numerical resolution is not enough to discriminate the behavior of the corrected spectrum for $k\sim 0$, which may be attained for analytic treatment if one generalizes the approach of the AdS/Ricci-flat correspondence of \cite{ads-rflat} to the presence of higher curvature terms. It is interesting to notice that the critical value of $k_c$ that may trigger the GL instability grows with the value of the Gauss-Bonnet coupling, namely the region of instability grows as the GB coupling is turned on, which can be seen in more detail in Figure 2. 

\begin{figure}[htb]
\begin{center}
\includegraphics[scale=0.49]{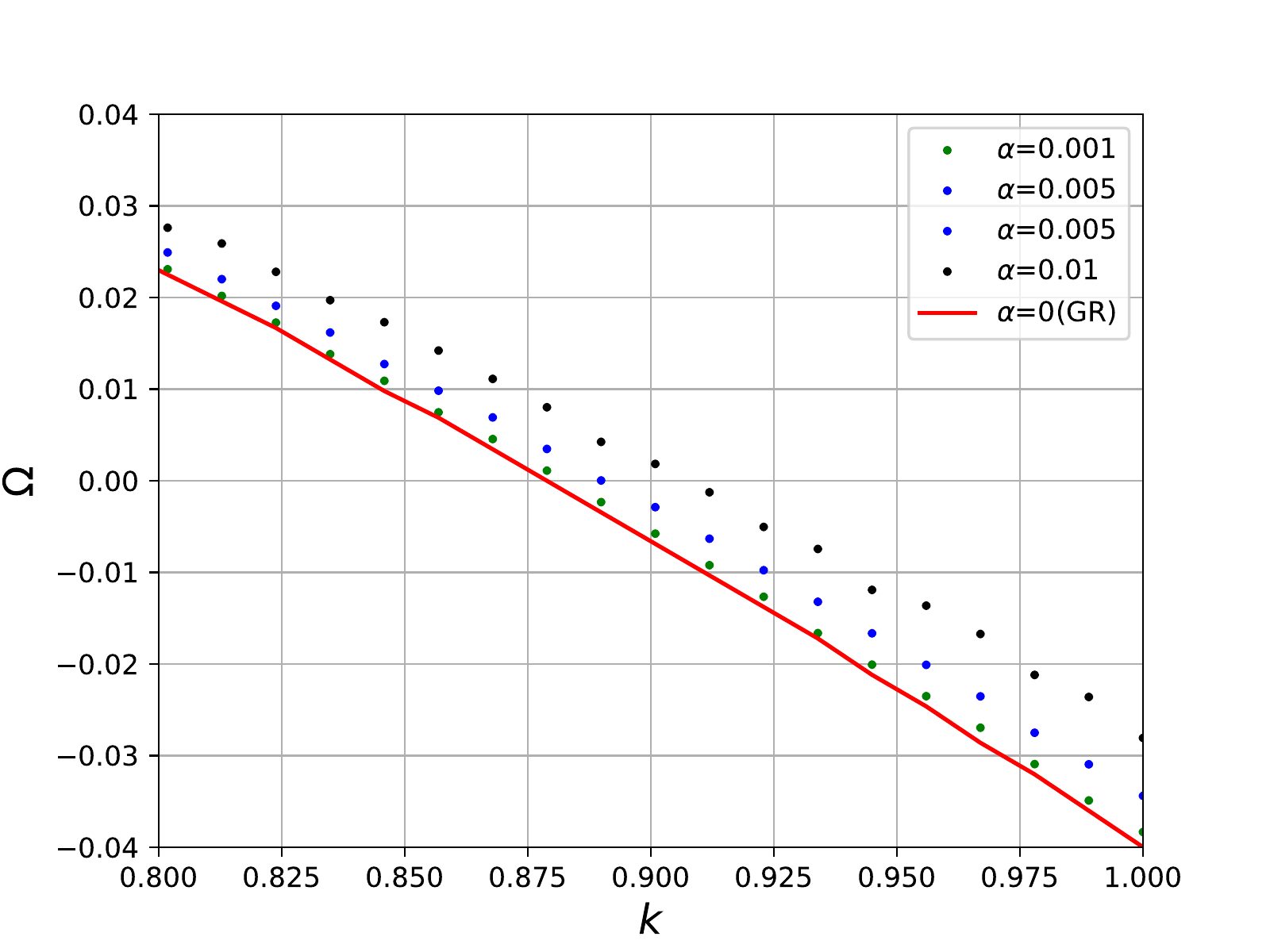}\quad\includegraphics[scale=0.49]{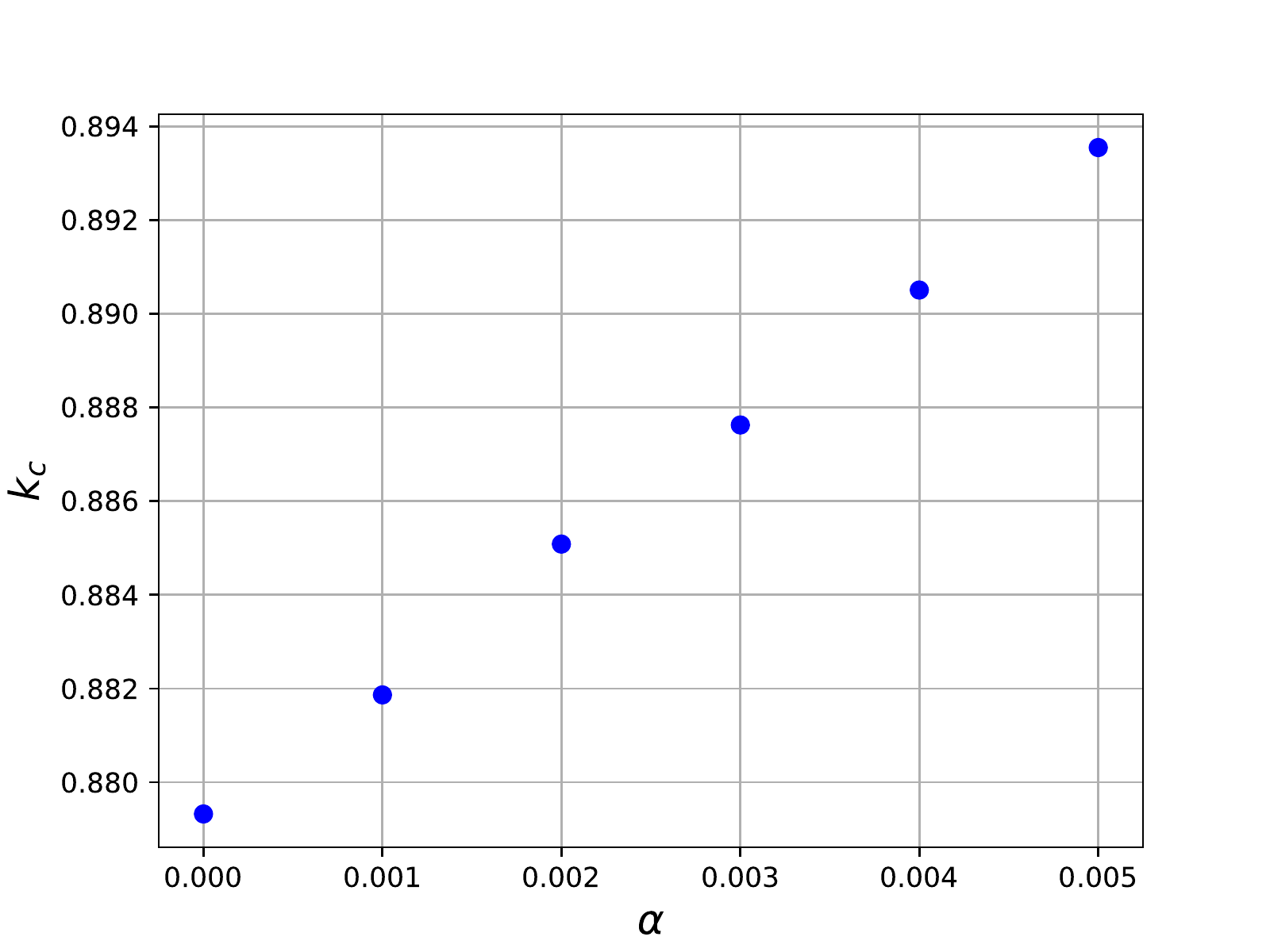}
\caption{Behavior of the critical wavelength that triggers the black string instability on the scalar mode, as a function of the $\alpha$ correction, and $r_+$=1.}
\end{center}
\end{figure}

\section{The $\alpha$-corrected, boosted black string}

A boosted black string, with momentum along the $z$ direction, can be obtained
by applying the transformation%
\begin{equation}\label{boost}
t\rightarrow\cosh\beta t+\sinh\beta z\text{ and }z\rightarrow\cosh\beta
z+\sinh\beta t\ ,
\end{equation}
to the metric \eqref{metricstring}.
This transformation corresponds to a boost with rapidity $v=-\tanh\beta$, and
it generates non-vanishing momentum along the $z$ direction, which can be
checked by computing the corresponding, $\alpha$-corrected ADM momentum, as in
GR \cite{Myers-Hovdebo-boosted-bs}. Since the new configuration is
characterized by a different set of global charges, it corresponds to a
physically different state in the phase space of the theory in spite of being
generated from the static solution  \eqref{metricstring} by a simple boost \eqref{boost}. Then, the $\alpha
$-corrected, boosted black string is
\begin{equation}\label{boosted}
ds_{\text{boosted}}^{2}=-b\left(  r\right)  dt^{2}+\frac{dr^{2}}{g\left(
r\right)  }+r^{2}d\sigma^{2}+b\left(  r\right)  dz^{2}+\left(  b\left(
r\right)  -f\left(  r\right)  \right)  \cosh^{2}\beta\left(  dt+\tanh\beta
dz\right)  ^{2}\ .
\end{equation}
Notice that the solution is asymptotically flat on a static frame, since
given $b,f$ and $g$ of equations \eqref{bs5df}-\eqref{bs5db}, the $g_{tz}$ component of the metric \eqref{boosted} vanishes as $r\rightarrow+\infty$.

Again, using the Iyer-Wald method, one can obtain the energy and the momentum of these
$\alpha$-corrected, boosted black strings, as conserved charges associated to
the Killing vectors $\partial_{t}$ and $\partial_{z}$, respectively. This yields
\begin{align}
Q\left(  \partial_{t}\right)    & =E=4\pi r_{+}\left(  \cosh^{2}%
\beta+1\right)  L_{z}+\frac{4\pi}{9r_{+}}\left(  47\cosh^{2}\beta-25\right)
L_{z}\alpha+\mathcal{O}\left(  \alpha^{2}\right)  \ ,\\
Q\left(  \partial_{z}\right)    & =P=-4\pi r_{+}\cosh\beta\sinh\beta L_{z}%
-\frac{188\pi}{9r_{+}}\cosh\beta\sinh\beta L_{z}\alpha+\mathcal{O}\left(
\alpha^{2}\right)  \ .
\end{align}
Going to Eddington-Finkelstein-like coordinates, one computes the surface gravity of the boosted black string, which leads to the following expression for the temperature
\begin{equation}
T=\frac{1}{4\pi r_+\cosh{\beta}}-\frac{11}{36\pi r_+^3\cosh{\beta}}\alpha+\mathcal{O}\left(
\alpha^{2}\right)\ .
\end{equation}
The entropy of the configuration is given by the Iyer-Wald formula
\begin{equation}
    S=\frac{1}{T}\int_{\mathcal{H}_+} \mathbf{Q}[\xi] \, ,
\end{equation}
where $\xi$ is the horizon generator. As expected, the expression for $\xi$ leads to the horizon velocity $v_h=-\tanh{\beta}$, and the entropy takes the form
\begin{equation}
    S=\cosh\beta\left(16 \pi^2 r_+^2 L_z+\frac{928 \pi^2}{9}L_z \alpha\right)+\mathcal{O}\left(
\alpha^{2}\right) \, .
\end{equation}
As before, one recovers the GR expression of reference \cite{Myers-Hovdebo-boosted-bs} when $\alpha$ goes to zero.

For black strings of a fixed length, the first law is fulfilled, namely
\begin{equation}
dM=TdS+v_hdP \ ,
\end{equation}
disregarding terms $\mathcal{O}\left(
\alpha^{2}\right)$. On a more general ensemble, one can also consider variations of $L_z$, as in \cite{Myers-Hovdebo-boosted-bs}. This leads to an extra work term in the first law, with the tension $\hat{\mathcal{T}}$ as the variable conjugate to $L_z$. In this case, the first law takes the form
\begin{equation}
dM=TdS+v_hdP+\hat{\mathcal{T}}dL_z \ ,
\end{equation}
where the tension acquires a correction with respect to that of black strings in GR, namely
\begin{equation}
\hat{\mathcal{T}}=4\pi r_+-\frac{100\pi}{9r_+}\alpha+\mathcal{O}\left(
\alpha^{2}\right).
\end{equation}

It is very interesting to notice that in contra-position to what occurs for boosted black strings in GR \cite{Kastor-Ray}, the inclusion of the higher curvature correction $\alpha$, spoils the validity of an standard Smarr Law, namely $2M$ is different from $3TS+\hat{\mathcal{T}}L_z+2v_hP$. As in the presence of a cosmological constant \cite{Kastor-Ray-first-law-boosted-BH, Dolan-BH-eqofstate-cosmoconstant}, one can restore the validity of a relation between finite thermodynamics quantities, i.e. the validity of the Euler relation of thermodynamics for a homogeneous system, by including in the first law a work term proportional to variations of the dimensionfull coupling constant $\alpha$. This approach leads to
\begin{equation}
dM=TdS+v_hdP+\hat{\mathcal{T}}dL_z+\mu_\alpha d\alpha \ ,
\end{equation}
with $\mu$ being the canonical conjugate of $\alpha$ and is defined as
\begin{equation}
\mu_\alpha=\frac{\partial M(S,P,L_z,\alpha)}{\partial\alpha}=-\frac{16 L_z\pi}{r_+} \ .
\end{equation}
With these expressions at hand, for the boosted black string characterized by the parameters $v_h$ and $r_+$, it is direct to prove the following Smarr-like formula
\begin{equation}
2M=3TS+\hat{\mathcal{T}}L_z+2v_hP+2\mu_\alpha \alpha \ ,
\end{equation}
which can also be obtained on dimensional grounds by a scaling argument. It would be interesting to have a physical understanding of the latter work term, which must be present if one insists on the validity of an Euler-like relation within this setup, and to explore its effect on the possible phase transitions that may be triggered by this new term. Notice also that the correction to the mass and the entropy are positive, while the contribution of the $\alpha$ term to the momentum of the boosted black string has the same sign as the uncorrected value\footnote{See the recent \cite{Newpapers-2-higherderivatvecorrectionstothermoofblackbranes, Newpapers-3-BHentropycorrectedfromstrings} for a string theory setup that leads to negative contributions to the entropy, due to $\alpha'$ corrections at fixed global charges, for black strings and black branes. This is in tension with standard expectations from the weak gravity conjecture.}.

\section{Further remarks}
In this paper, we computed the effect of the Gauss-Bonnet term on the spectrum of the scalar mode that triggers the GL instability in a regime in which the Gauss-Bonnet coupling can be treated as a perturbation, which is necessary if we want to interpret this $R^2$ term as a higher curvature correction coming from string theory. Recently, the static, spherically symmetric black hole solution of this setup was obtained for arbitrary dimensions including the dilaton in a frame that leads to second order field equations \cite{Agurto-Oliva-alpha} (see also \cite{Callan-Myers-bhs-in-st} for the original computation on the frame leading to fourth order field equations). Furthermore, the mentioned second order frame \cite{Maeda-Field-Redef} allowed to identify the four-dimensional, rotating solution containing terms of order $\alpha,\ a, \ a^2$ and $\alpha a$. It would be interesting to explore effects of the rotation of the four-dimensional metric on the black string and evaluate the interplay between the superradiant instability and the GL instability as in \cite{Newpapers-superradiance-1, Newpapers-0-superradiance}. The presence of $a^2$ terms of the perturbative solution constructed in \cite{Agurto-Oliva-alpha} allows the existence of an ergoregion and therefore a potential superradiant behavior. As with the dilaton, it is also known that the presence of fluxes and scalars with non-minimal couplings, permits the construction of closed form solutions of homogeneous black strings and black branes in presence of higher curvature terms \cite{Cisterna-Oliva-Fuenzalida-2, Cisterna-Mora-1-BSs-in-Lovelock, Cisterna-and-beyond, Cisterna-Corral-del-Pino-5}. The effect of such terms on the GL instability still remains an open problem, but it is important to mention that some of the field theories that involve non-minimally coupled scalar fields may also admit strongly hyperbolic formulations as shown in \cite{Reall-Kovacs-Stronghyperbolicity}.

As a simplified setup to evaluate the effect of the higher curvature corrections, one can consider the regime in which the higher curvature terms completely dominate over GR terms, which may be consistent if one goes beyond the perturbative regime. This approach was considered in pure $R^2$ \cite{Giacomini-1, Giacomini-3} and pure $R^3$ \cite{Giacomini-2} Lovelock theories, exploiting the fact that these theories admit exact, homogeneous black strings \cite{Giribet-Oliva-Troncoso, Kastor-Mann}. In these works it was shown that the GL instability persists, but in this regime the region of instability shrinks as one goes from $R$ to $R^2$ and then to the $R^3$ theory. Recently, in the context of the Large-D expansion, these results were recovered analytically for the $R^2$ case \cite{Newpapers-4-PhasesofBSatlargeD} and also by the inclusion NLO terms in the $1/D$ correction it was shown that the critical dimension increases with the value of the coupling (see also \cite{Suzuki-2}).

Finally, it is important to mention that in the context of M-theory, still at a perturbative level, the authors of \cite{Hyakutake-M} considered the $R^4$ correction in the analysis of the thermodynamic analogue of the GL instability of boosted black strings. More recently, the authors of \cite{Newpapers-1-newBSwithapcorrections} showed that the singularity of a two-dimensional black hole can be smoothed out by using a recent classification of the higher curvature corrections from a Bottom-Up approach via T-duality \cite{Hohm-zwie-1, Hohm-zwie-2, Hohm-zwie-3}. Furthermore, they elaborated on the regularity of the corresponding black string constructed out from this black hole. This setup is simple enough as to study the dynamics of black strings in three dimensions, considering higher curvature corrections of arbitrary high power.

\begin{acknowledgments}
We thank Tomás Andrade, Adolfo Cisterna, Nicolás Grandi, Gaston Giribet, Rodrigo Olea and Ricardo Stuardo for useful discussions and insight. C. H. is partially funded by the National Agency for Research and Development ANID -
PAI Grant No. 77190078 and also thanks the support of FONDECYT Grant No. 1181047.
 J.O. is partially supported by FONDECYT grants 1221504 and 1210635 and by Proyecto de cooperaci\'{o}n internacional 2019/13231-7
FAPESP/ANID. The work of M.O. is partially funded by Beca ANID de Doctorado 21222264.
\end{acknowledgments}

\end{document}